\pdfoutput=1
\documentclass[10pt]{article}
\usepackage{graphicx}

\def\Title#1{\begin{center} {\Large #1 } \end{center}}
\def\Author#1{\begin{center}{ \sc #1} \end{center}}
\def\Address#1{\begin{center}{ \it #1} \end{center}}

\newcommand\pubblock{\rightline{\begin{tabular}{l} Proceedings of the Second Annual LHCP\\ \pubnumber\\
         \pubdate  \end{tabular}}}

\newenvironment{Abstract}{\begin{quotation} \begin{center} 
             \large ABSTRACT \end{center}\bigskip 
      \begin{center}\begin{large}}{\end{large}\end{center} \end{quotation}}

\newenvironment{Presented}{\begin{quotation} \begin{center} 
             PRESENTED AT\end{center}\bigskip 
      \begin{center}\begin{large}}{\end{large}\end{center} \end{quotation}}

\def\beq{\begin{equation}}
\def\eeq#1{\label{#1}\end{equation}}
\def\eeqn{\end{equation}}

\def\beqa{\begin{eqnarray}}
\def\eeqa#1{\label{#1}\end{eqnarray}}
\def\eeqan{\end{eqnarray}}

\let\bar=\overbar

\def\D{{\cal D}}

\def\Dslash{\not{\hbox{\kern-4pt $D$}}}
\def\dslash{\not{\hbox{\kern-2pt $\del$}}}

\def\msb{{\bar{\ssstyle M \kern -1pt S}}}

\usepackage{xspace}

\usepackage{amsmath} 
\usepackage{amssymb}                                                                                                                      
\usepackage{ifthen} 
\usepackage{lineno}
\usepackage{amsfonts}
\usepackage{upgreek} 

\newboolean{uprightparticles}
\setboolean{uprightparticles}{false} 

\def\lhcb {\mbox{LHCb}\xspace}
\def\ux85 {\mbox{UX85}\xspace}

\ifthenelse{\boolean{uprightparticles}}%
{

 \def\Pmu         {\ensuremath{\upmu}\xspace}

 \def\Ppi         {\ensuremath{\uppi}\xspace}

 \def\PDelta      {\ensuremath{\Delta}\xspace}                 
 \def\PXi      {\ensuremath{\Xi}\xspace}                 
 \def\PLambda      {\ensuremath{\Lambda}\xspace}                 
 \def\PSigma      {\ensuremath{\Sigma}\xspace}                 
 \def\POmega      {\ensuremath{\Omega}\xspace}                 
 \def\PUpsilon      {\ensuremath{\Upsilon}\xspace}                 
 
 \def\PB      {\ensuremath{\mathrm{B}}\xspace}                 
                  
 \def\PD      {\ensuremath{\mathrm{D}}\xspace}

 \def\PK      {\ensuremath{\mathrm{K}}\xspace}

 \def\Pc      {\ensuremath{\mathrm{c}}\xspace}                 
                  
 \def\Pe      {\ensuremath{\mathrm{e}}\xspace}

 \def\Pi      {\ensuremath{\mathrm{i}}\xspace}

 \def\Ps      {\ensuremath{\mathrm{s}}\xspace}

}
{

 \def\Pmu         {\ensuremath{\mu}\xspace}

 \def\Ppi         {\ensuremath{\pi}\xspace}

 \mathchardef\PDelta="7101
 \mathchardef\PXi="7104
 \mathchardef\PLambda="7103
 \mathchardef\PSigma="7106
 \mathchardef\POmega="710A
 \mathchardef\PUpsilon="7107
                  
 \def\PB      {\ensuremath{B}\xspace}                 
                  
 \def\PD      {\ensuremath{D}\xspace}

 \def\PK      {\ensuremath{K}\xspace}

 \def\Pc      {\ensuremath{c}\xspace}                 
                  
 \def\Pe      {\ensuremath{e}\xspace}

 \def\Pi      {\ensuremath{i}\xspace}

 \def\Ps      {\ensuremath{s}\xspace}

}

\def\en         {\ensuremath{\Pe^-}\xspace}   
\def\ep         {\ensuremath{\Pe^+}\xspace}

\def\mup        {\ensuremath{\Pmu^+}\xspace}
\def\mun        {\ensuremath{\Pmu^-}\xspace} 
\def\mumu       {\ensuremath{\Pmu^+\Pmu^-}\xspace}

\def\ellell     {\ensuremath{\ell^+ \ell^-}\xspace}

\def\squark    {\ensuremath{\Ps}\xspace}

\def\cquark    {\ensuremath{\Pc}\xspace}
\def\cquarkbar {\ensuremath{\overline \cquark}\xspace}
\def\ccbar     {\ensuremath{\cquark\cquarkbar}\xspace}

\def\pion  {\ensuremath{\Ppi}\xspace}

\def\pip   {\ensuremath{\pion^+}\xspace}
\def\pim   {\ensuremath{\pion^-}\xspace}

\def\kaon  {\ensuremath{\PK}\xspace}
  \def\Kbar  {\kern 0.2em\overline{\kern -0.2em \PK}{}\xspace}

\def\Kz    {\ensuremath{\kaon^0}\xspace}
\def\Kzb   {\ensuremath{\Kbar^0}\xspace}
\def\KzKzb {\ensuremath{\Kz \kern -0.16em \Kzb}\xspace}
\def\Kp    {\ensuremath{\kaon^+}\xspace}
\def\Km    {\ensuremath{\kaon^-}\xspace}

\def\KpKm  {\ensuremath{\Kp \kern -0.16em \Km}\xspace}

\def\Kstarz  {\ensuremath{\kaon^{*0}}\xspace}

\def\Kstarp  {\ensuremath{\kaon^{*+}}\xspace}

  \def\Dbar    {\kern 0.2em\overline{\kern -0.2em \PD}{}\xspace}
\def\D       {\ensuremath{\PD}\xspace}

\def\Dz      {\ensuremath{\D^0}\xspace}
\def\Dzb     {\ensuremath{\Dbar^0}\xspace}
\def\DzDzb   {\ensuremath{\Dz {\kern -0.16em \Dzb}}\xspace}
\def\Dp      {\ensuremath{\D^+}\xspace}
\def\Dm      {\ensuremath{\D^-}\xspace}

\def\DpDm    {\ensuremath{\Dp {\kern -0.16em \Dm}}\xspace}

\def\B       {\ensuremath{\PB}\xspace}
  \def\Bbar    {\kern 0.18em\overline{\kern -0.18em \PB}{}\xspace}

\def\Bz      {\ensuremath{\B^0}\xspace}
\def\Bzb     {\ensuremath{\Bbar^0}\xspace}
\def\Bu      {\ensuremath{\B^+}\xspace}

\def\Bp      {\ensuremath{\Bu}\xspace}

\def\Bs      {\ensuremath{\B^0_\squark}\xspace}

  \def\Y#1S{\ensuremath{\PUpsilon{(#1S)}}\xspace}

\def\Lbar {\ensuremath{\kern 0.1em\overline{\kern -0.1em\Lambda\kern -0.05em}\kern 0.05em{}}\xspace}

\def\BF         {{\ensuremath{\cal B}\xspace}}

\newcommand{\decay}[2]{\ensuremath{#1\!\to #2}\xspace}         

\def\to                 {\ensuremath{\rightarrow}\xspace}

\def\qsq       {\ensuremath{q^2}\xspace}

\def\AT#1     {\ensuremath{A_{\mathrm{T}}^{#1}}\xspace}           

\def\C#1      {\ensuremath{\mathcal{C}_{#1}}\xspace}                       
\def\Cp#1     {\ensuremath{\mathcal{C}_{#1}^{'}}\xspace}                    
\def\Ceff#1   {\ensuremath{\mathcal{C}_{#1}^{\mathrm{(eff)}}}\xspace}        
\def\Cpeff#1  {\ensuremath{\mathcal{C}_{#1}^{'\mathrm{(eff)}}}\xspace}       
\def\Ope#1    {\ensuremath{\mathcal{O}_{#1}}\xspace}                       
\def\Opep#1   {\ensuremath{\mathcal{O}_{#1}^{'}}\xspace}                    

\newcommand{\tev}{\ensuremath{\mathrm{\,Te\kern -0.1em V}}\xspace}
\newcommand{\gev}{\ensuremath{\mathrm{\,Ge\kern -0.1em V}}\xspace}
\newcommand{\mev}{\ensuremath{\mathrm{\,Me\kern -0.1em V}}\xspace}
\newcommand{\kev}{\ensuremath{\mathrm{\,ke\kern -0.1em V}}\xspace}
\newcommand{\ev}{\ensuremath{\mathrm{\,e\kern -0.1em V}}\xspace}
\newcommand{\gevc}{\ensuremath{{\mathrm{\,Ge\kern -0.1em V\!/}c}}\xspace}
\newcommand{\mevc}{\ensuremath{{\mathrm{\,Me\kern -0.1em V\!/}c}}\xspace}
\newcommand{\gevcc}{\ensuremath{{\mathrm{\,Ge\kern -0.1em V\!/}c^2}}\xspace}
\newcommand{\gevgevcccc}{\ensuremath{{\mathrm{\,Ge\kern -0.1em V^2\!/}c^4}}\xspace}
\newcommand{\mevcc}{\ensuremath{{\mathrm{\,Me\kern -0.1em V\!/}c^2}}\xspace}

\def\invfb   {\ensuremath{\mbox{\,fb}^{-1}}\xspace}

\def\deriv {\ensuremath{\mathrm{d}}}

\def\gsim{{~\raise.15em\hbox{$>$}\kern-.85em
          \lower.35em\hbox{$\sim$}~}\xspace}
\def\lsim{{~\raise.15em\hbox{$<$}\kern-.85em
          \lower.35em\hbox{$\sim$}~}\xspace}

\def\tell1  {TELL1\xspace}
\def\ukl1   {UKL1\xspace}

\newcommand\pubnumber{}

\newcommand\pubdate{\today}

\def\affiliation{
On behalf of the LHCb Experiment, \\
Department of Physics, \\
Wariwck University, Coventry, UK }

\def\support{\footnote{Work supported by  the Royal Society}}

\begin{document}

\large
\begin{titlepage}
\pubblock

\vfill
\Title{ Rare FCNC top, beauty and charm decays }
\vfill

\Author{Thomas Blake \support}
\Address{\affiliation}
\vfill
\begin{Abstract}

Rare flavour changing neutral current (FCNC) decays of top, beauty and charm quarks can provide a powerful probe for as yet unobserved particles. Recent results on FCNC $b \to s$, $c \to u$ and $t$ transitions from the LHC experiments are reviewed. Particular attention is paid to the angular distribution of the \decay{\Bz}{\Kstarz\mumu} decay, where a measurement performed by LHCb shows a local discrepancy of 3.7 standard deviations with respect to the SM prediction. Using the decay \decay{\Bp}{\Kp\pim\pip\gamma}, LHCb have also been able to demonstrate the polarisation of photons  produced in \decay{b}{s\gamma} transitions. More work is needed both experimentally and theoretically to understand if the Standard Model description of these rare FCNC processes is correct. 
\end{Abstract}
\vfill

\begin{Presented}
The Second Annual Conference\\
 on Large Hadron Collider Physics \\
Columbia University, New York, U.S.A \\ 
June 2-7, 2014
\end{Presented}
\vfill
\end{titlepage}
\def\thefootnote{\fnsymbol{footnote}}
\setcounter{footnote}{0}
%

\normalsize 


\section{Introduction}

In the Standard Model (SM), the only flavour changing interaction is the charged current interaction. Flavour changing neutral current (FCNC) $b \to s (d)$, $c \to u$ and $t \to c (u)$ transitions can therefore only occur at loop order in the SM. This makes these processes rare. The lack of a dominant tree-level SM process also makes these FCNC sensitive to contributions from new, as yet unobserved, particles that can enter in competing loop order diagrams. The presence of these new particles can show up either as an increase or, due to interference, decrease in the rate of particular decays or as a change in the angular distribution of the particles in the detector.

\section{Leptonic decays}
 
The leptonic decays \decay{\Bz}{\mumu} and \decay{\Bs}{\mumu} are extremely rare in the SM. They are both loop and helicity suppressed. This leads to an expected branching fraction of ~\cite{Bobeth:2013uxa}
 
\begin{equation}
\BF(\decay{\Bs}{\mumu}) = (3.65\pm 0.23)\times 10^{-9} ~~,
\end{equation}  
 
\noindent where the largest uncertainties are from the \Bs decay constant (predicted with an uncertainty of 4\% from Lattice QCD calculations~\cite{Aoki:2013ldr}) and the CKM matrix elements. The rate of the \Bz decay is further suppressed in the SM by the ratio of the CKM elements $|V_{td}/V_{ts}|^{2}$.  In models with an extended Higgs sector, the helicity suppression can be lifted and the branching fraction of the decays significantly increased. 

In 2013, the CMS and \lhcb experiments both reported evidence for the \decay{\Bs}{\mumu} decay at the level of $\sim4$ standard deviations~\cite{Chatrchyan:2013bka, Aaij:2013aka}. Key to both analyses was their ability to reject backgrounds consisting of random combinations of muons from $b$-hadron decays or $b$-hadron decays where one or more particles was mistakenly identified as a muon. No evidence for the \Bz decay was seen by either experiment. A na\"ive combination of the measurements by CMS and \lhcb  yields~\cite{combination:conf}  
 
\begin{equation}
\begin{split}
\BF(\decay{\Bs}{\mumu}) &= ( 2.9\pm 0.7 ) \times 10^{-9} \\
\BF(\decay{\Bz}{\mumu}) &= ( 3.6^{\,+1.6}_{\,-1.4} ) \times 10^{-10} ~. \\
\end{split} 
\end{equation} 

\noindent  These averages are compatible with the SM expectation, setting constraints on contributions of new virtual particles. 
 

\begin{figure}[htb]
\centering
\begin{minipage}[c]{0.48\textwidth}
\includegraphics[width=\linewidth]{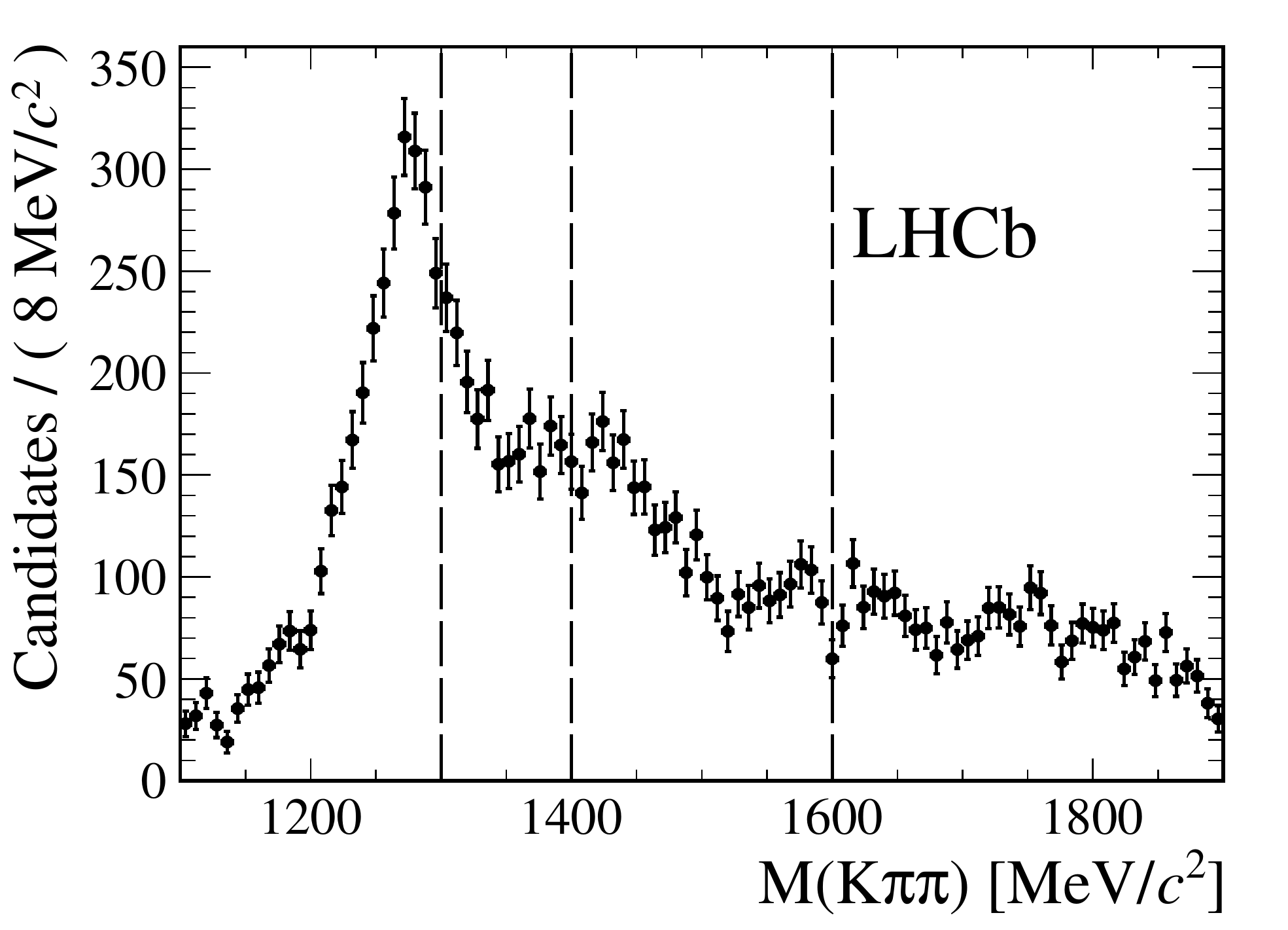} 
\end{minipage}
\begin{minipage}[c]{0.48\textwidth}
\includegraphics[width=\linewidth]{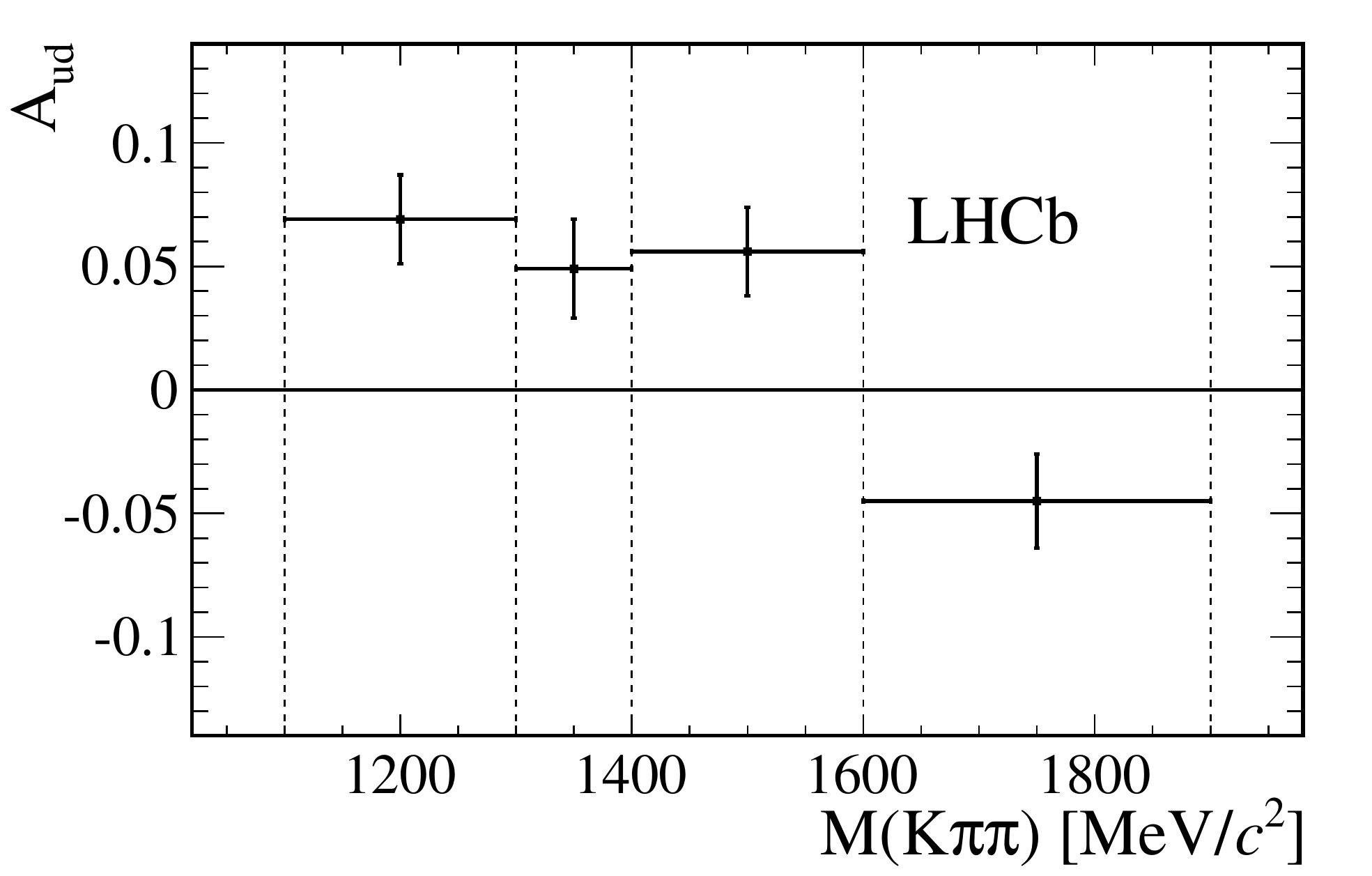} 
\end{minipage}
\caption{
Invariant $\Kp\pim\pip$ mass distribution of selected \decay{\Bp}{\Kp\pim\pip\gamma} candidates (left) and the up-down asymmetry of the photon with respect to the $\Kp\pim\pip$ system in four bins of $\Kp\pim\pip$ mass (right).
\label{fig:updown}
}
\end{figure}

\section{Photon polarisation} 

In the SM, photons produced in \decay{b}{s\gamma} transitions are expected to be almost purely left-hand polarised due to the coupling of the $b$- and $s$-quark to a virtual $W$ boson. Neglecting QCD contributions, the right-handed component ($b_{L} \to \gamma_{R} s_{R}$) is suppressed by the ratio of the $s$- to the $b$-quark mass, making it vanishingly small. In many extensions of the SM, the photon is produced unpolarised because there is no preferred left- or right-handed coupling. 

One way to test the photon polarisation is using the photon direction with respect to the $\Kp\pim\pip$  system in \decay{\Bp}{\Kp\pim\pip\gamma} decays~\cite{Gronau:2002rz}. The decay distribution can be written as 

\begin{equation}
\frac{1}{\Gamma}\frac{\deriv\Gamma}{\deriv\cos\theta} = \sum_{n = 0, 2, 4} a_n  \cos^{n} \theta +  \lambda \sum_{n = 1, 3} a_n \cos^{n} \theta ~~,
\end{equation}

\noindent where $\theta$ is the angle between the direction of the photon and the plane defined by the \pip and \pim and the coefficient $a_n$ depends on the resonant structure of the $\Kp\pim\pip$, $\Kp\pim$ and $\pip\pim$ systems. The up-down asymmetry of the photon with respect to this plane is proportional to the photon-polarisation, $\lambda$. 

The LHCb experiment has performed a first measurement of this up-down asymmetry using a dataset corresponding to 3\invfb of integrated luminosity~\cite{Aaij:2014wgo}. The complete dataset contains $13876 \pm153$ signal candidates. It is split into four regions of $\Kp\pim\pip$ invariant mass, shown in Fig.~\ref{fig:updown}, designed to separate different resonant contributions. The measured up-down asymmetry in the four regions is also shown in Fig.~\ref{fig:updown}. Combining the four regions, evidence for non-zero photon polarisation is observed at the level of $5.2\sigma$. This is the first observation of photon polarisation in radiative $b$-hadron decays. Whilst clear evidence for polarisation is seen, an understanding of the hadronic system is needed to compare the measured asymmetry to the left-hand polarisation prediction of the SM. Work is needed on the experimental side to understand the different resonant contributions to the $\Kp\pim\pip$ system and on the theoretical side to convert the measured $\Kp\pim\pip$ spectrum and the up-down asymmetry into a measurement of the photon polarisation. 

\section{Angular distribution of \decay{\Bz}{\Kstarz\mumu}} 

The polarisation of virtual photons can also be probed using the angular distribution of \decay{\Bz}{\Kstarz\mumu} decays, where \decay{\Kstarz}{\Kp\pim}, at low dimuon invariant mass squared, \qsq.  The angular distribution of this decay can be defined by three angles, $\theta_{\ell}$, $\theta_{K}$ and $\phi$, as (see for example Ref.~\cite{Altmannshofer:2008dz})

\begin{equation}
\frac{\deriv^{4}\Gamma}{\deriv\cos\theta_{\ell}\,\deriv\cos\theta_{K}\,\deriv\phi\,\deriv\qsq} = \frac{9}{32\pi} \sum_{i} J_{i}(\qsq) f_{i}(\cos\theta_{\ell},\cos\theta_{K},\phi).
\end{equation} 

\noindent Here, $\theta_{\ell}$ is defined by the direction of the \mup (\mun) with respect to the \Bz (\Bzb) in the dimuon rest frame and $\theta_{K}$ by the direction of the kaon with respect to the \Bz (\Bzb). The angle $\phi$ is the angle between the plane containing the \mup and \mun and the plane containing the kaon and pion. The angular distribution can be particularly sensitive to the contribution of new virtual particles through their interference with the SM contributions.  Different angular terms, $J_{i}(\qsq)$, are sensitive to different \Kstarz polarisation states and provide complementary information. 

\begin{figure}[htb]
\centering
\includegraphics[width=0.48\textwidth]{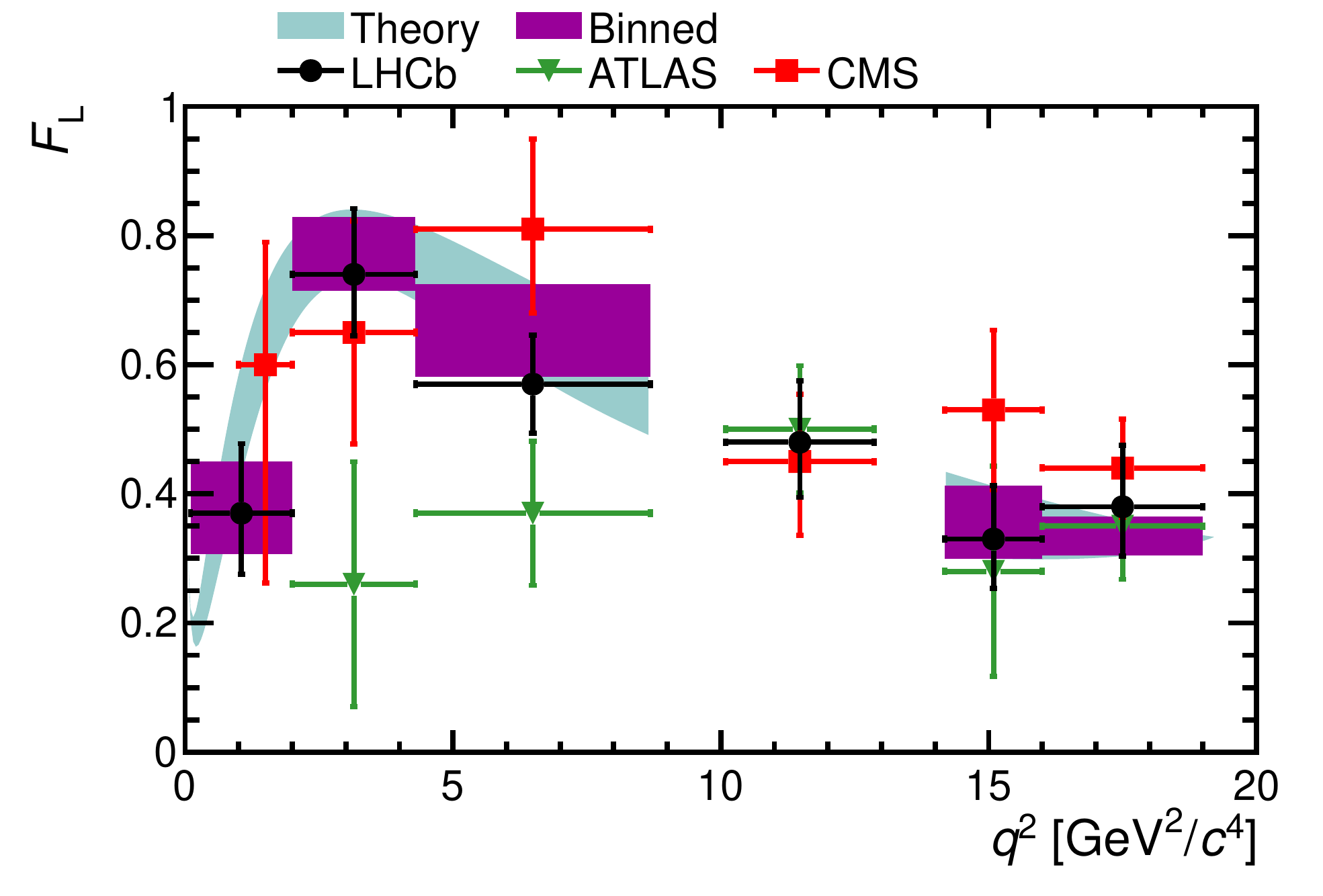} 
\includegraphics[width=0.48\textwidth]{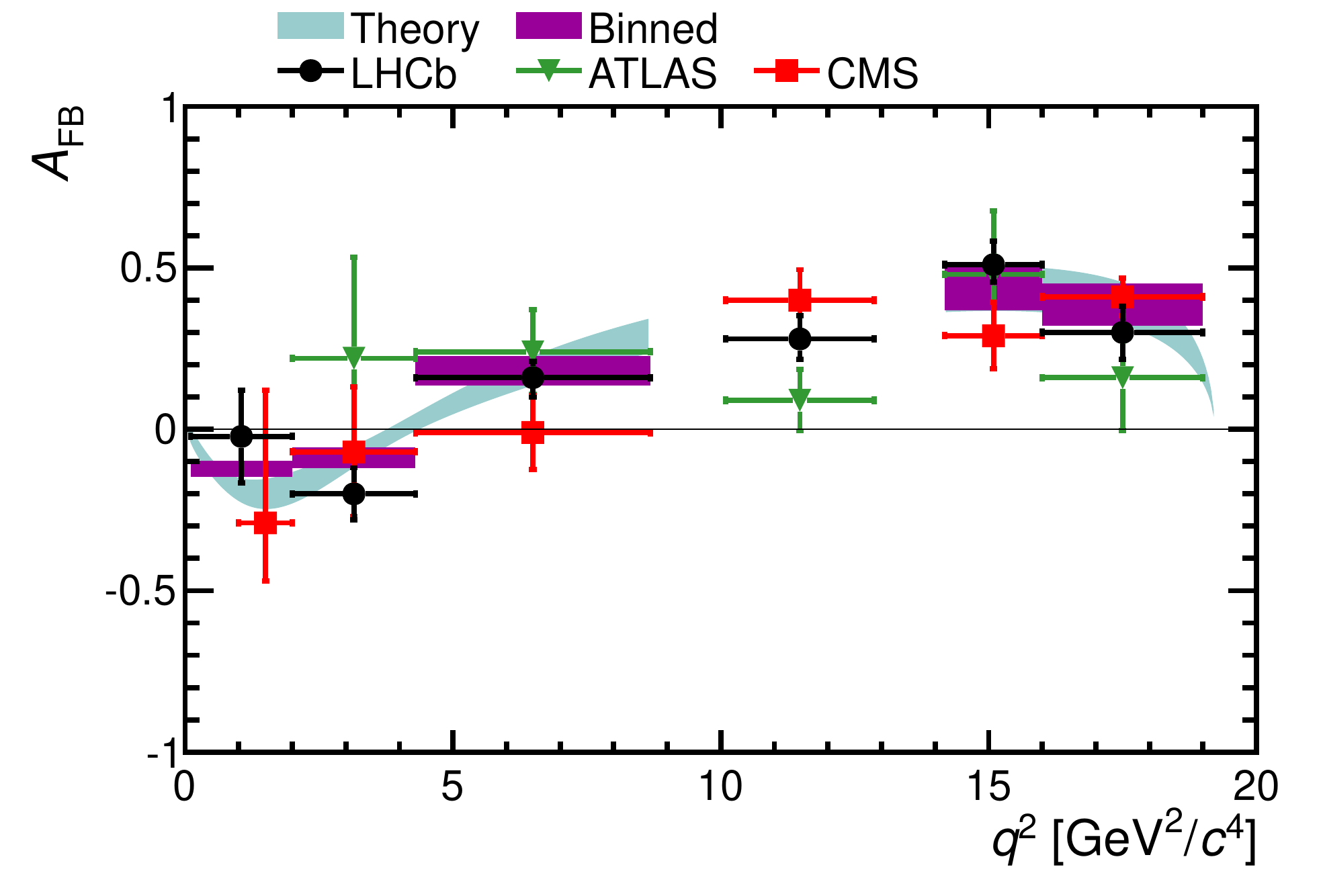}  \\ 
\caption{Fraction of longitudinal polarisation, $F_{\rm L}$, of the \Kstarz produced in \decay{\Bz}{\Kstarz\mumu} decays (left) and forward-backward asymmetry, $A_{\rm FB}$, of the dimuon system (right) as a function of the dimuon invariant mass squared, \qsq. Results of fits to the datasets collected by the ATLAS~\cite{ATLAS:KstarMuMu}, CMS~\cite{Chatrchyan:2013cda} and LHCb~\cite{Aaij:2013iag} experiments are included. The data are overlaid with a SM prediction based on Ref.~\cite{Bobeth:2011gi} and references therein.} 
\label{fig:angular}
\end{figure}

The ATLAS, CMS and \lhcb experiments have all performed measurements of the \decay{\Bz}{\Kstarz\mumu} angular distribution using the data they collected in 2011~\cite{ATLAS:KstarMuMu,Chatrchyan:2013cda,Aaij:2013iag}. Due to the rarity of the decay and the small number of candidates that are reconstructed, the experiments do not simultaneously fit for all of the angular terms. ATLAS and CMS measure only the longitudinal polarisation fraction of the \Kstarz and the forward-backward asymmetry of the dimuon system, $A_{\rm FB}$, shown in Fig.~\ref{fig:angular}. The \lhcb experiment also measures the asymmetry between the two transverse \Kstarz polarisations, which is particularly sensitive to the handedness of the photon polarisation. All of these measurements are consistent with SM expectations.

\begin{figure}[htb]
\centering
\includegraphics[width=0.48\textwidth]{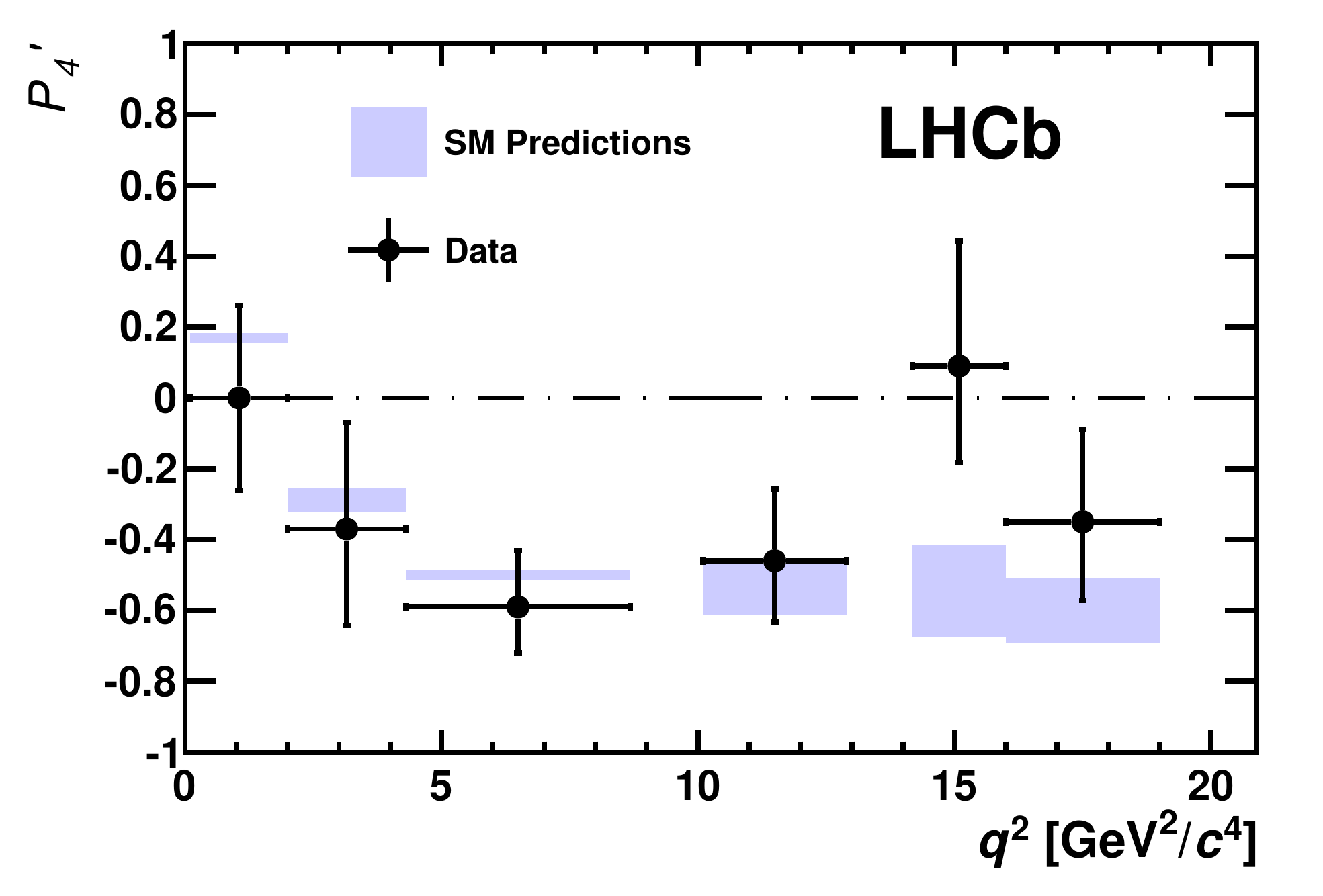} 
\includegraphics[width=0.48\textwidth]{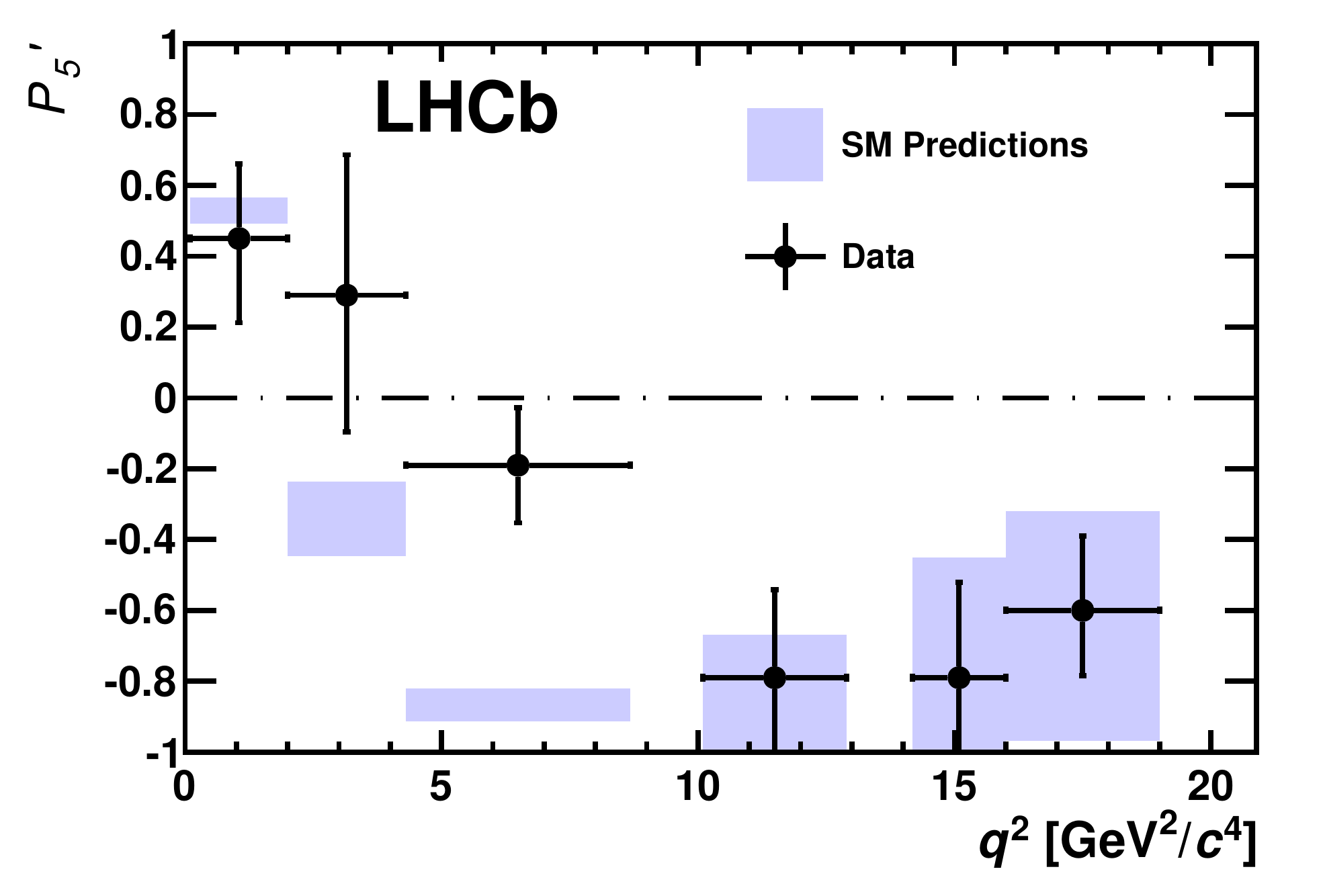}  \\ 
\caption{Form-factor free observables $P'_4$ and $P'_5$, measured by the LHCb collaboration~\cite{Aaij:2013qta} in \decay{\Bz}{\Kstarz\mumu} decays in bins of dimuon invariant mass squared, \qsq. The data are overlaid with a SM prediction described in Ref.~\cite{Descotes-Genon:2013vna}.} 
\label{fig:p5prime}
\end{figure}

The \lhcb experiment has also made first measurements of two new observables~\cite{Aaij:2013qta}, $P_{4,5}'$, which are free from form-factor uncertainties at leading order~\cite{Descotes-Genon:2013vna}. The result of these measurements is shown in Fig.~\ref{fig:p5prime}. Interestingly,  a large local discrepancy of $3.7\sigma$ is seen between the measurement and the SM prediction in the \qsq range $4.3 < \qsq < 8.68\gev^{2}/c^{4}$ for $P'_5$. Several recent attempts have been made to understand this anomaly by performing global fits to the measurements of the angular distribution made by the LHC experiments, CDF and the B factories. The data are best described by a model in which a new vector current is introduced that destructively interferes with the SM contributions~\cite{Descotes-Genon:2013wba,Altmannshofer:2013foa,Beaujean:2013soa}. This is most visible at low \qsq due to additional interference with the virtual photon contribution in the SM.  A new vector current is best explained in models that introduce a $Z'$ boson with flavour violating couplings~\cite{Gauld:2013qja}. It is less easy to explain this type of deviation in supersymmetric models. It is also possible that part of this discrepancy can be explained by an underestimate of the uncertainty on the SM prediction, coming from the treatment of the form-factors~\cite{Jager:2012uw} or from our understanding of \ccbar contributions to the decay~\cite{Lyon:2014hpa}. The datasets collected by the experiments in 2012 may help to shed light on the situation.  

\newpage

\section{Branching fraction of semileptonic \decay{b}{s\mumu} decays} 

If there is a new vector current that explains the anomaly in the \decay{\Bz}{\Kstarz\mumu} angular distribution, then its influence should also show up in other decays involving \decay{b}{s\mumu} quark level transitions. In particular, the destructive interference will result in branching fractions that are below their SM expectation. 

\begin{figure}[htb]
\centering
\begin{minipage}[c]{0.48\textwidth}
\includegraphics[width=\linewidth]{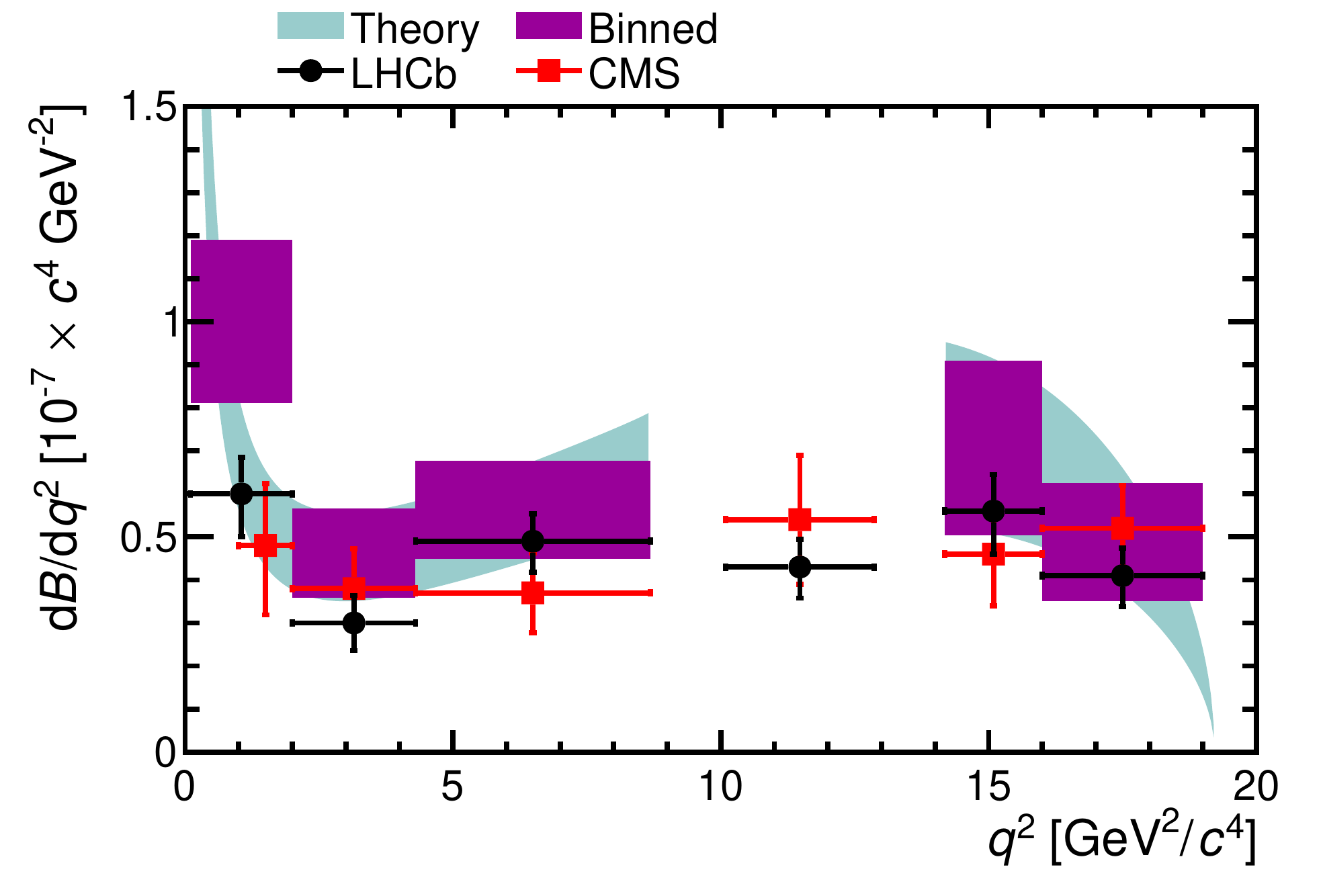} 
\end{minipage}
\begin{minipage}[c]{0.48\textwidth}
\includegraphics[width=\linewidth]{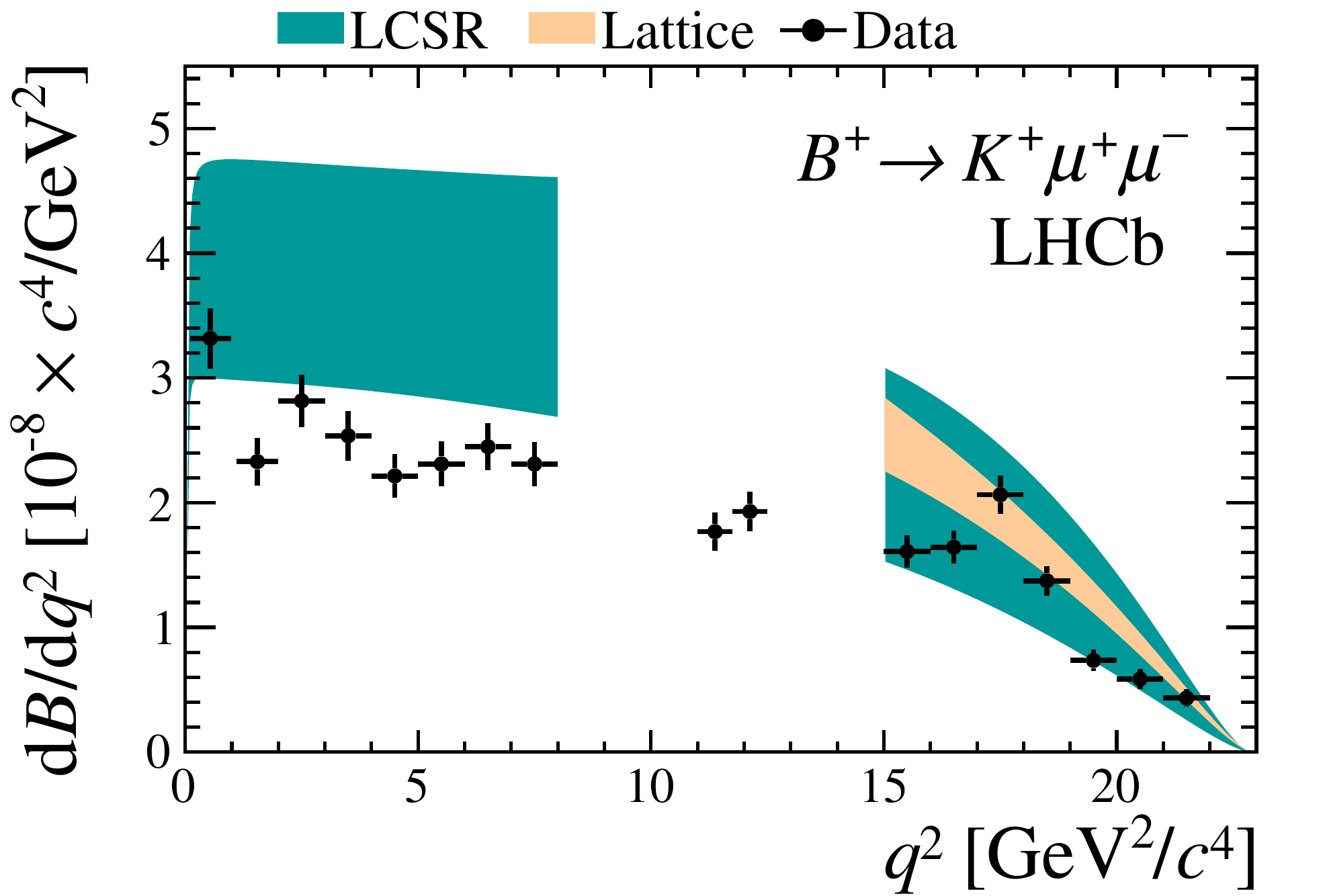} 
\end{minipage}
\caption{Differential branching fraction of the decay \decay{\Bz}{\Kstarz\mumu} measured by the CMS and LHCb experiments (left) and of the decay \decay{\Bp}{\Kp\mumu} measured by the LHCb experiment (right) as a function of the dimuon invariant mass squared, \qsq. The data are overlaid with SM predictions described in Refs.~\cite{Bobeth:2011gi}\cite{Bobeth:2011nj}. A prediction for the differential branching fraction of the \decay{\Bp}{\Kp\mumu} decay using form-factors from lattice~\cite{Bouchard:2013mia} is also included. 
}. 
\label{fig:decayrate}
\end{figure}

In Refs.~\cite{Aaij:2013iag,Aaij:2014pli}, the \lhcb collaboration performed measurements of the differential branching fraction of the decays \decay{\Bz}{\Kstarz\mumu}, \decay{\Bp}{\Kstarp\mumu}, \decay{\Bz}{\Kz\mumu} and \decay{\Bp}{\Kp\mumu}. The measurements for  \decay{\Bz}{\Kstarz\mumu} and \decay{\Bp}{\Kp\mumu} are shown in Fig.~\ref{fig:decayrate}. All of these measurements are below their SM expectation. However, they are consistent with the SM when accounting for the large uncertainties on the SM predictions coming from the $B \to K^{(*)}$ form-factors.

\section{Lepton universality} 

The large form-factor uncertainties in \decay{B}{K^{(*)}\mumu} decays can be cancelled by forming appropriate ratios or asymmetries between similar decays. An example is the ratio

\begin{equation}
R_{\rm K} = \frac{\int_{1\gev^{2}/c^{4}}^{6\gev^{2}/c^{4}} \frac{\deriv\Gamma[\decay{\Bp}{\Kp\mumu}]}{\deriv\qsq} \deriv\qsq}{\int_{1\gev^{2}/c^{4}}^{6\gev^{2}/c^{4}} \frac{\deriv\Gamma[\decay{\Bp}{\Kp\ep\en}]}{\deriv\qsq} \deriv\qsq}~.
\end{equation}

\noindent This ratio is very close to unity in the SM, due to the universality of the electroweak couplings to leptons. It differs from unity in the SM only because of tiny contributions from Higgs penguins and small phase-space differences, which amount to a correction of $\mathcal{O}(10^{-3})$ . The $B \to K$ form-factors cancel completely in this ratio. In models with an extended Higgs sector, $R_{\rm K}$ can be larger than unity because $m(\mu) > m(e)$. In models with an additional $Z'$ boson, $R_{\rm K}$ can be both larger or smaller than unity depending on how the $Z'$ couples to the different flavours of lepton.

The LHCb experiment has recently produced the most precise measurement of $R_{\rm K}$ to date~\cite{LHCb:RK} of  

\begin{equation}
R_{\rm K} (1 < \qsq < 6\gev^{2}/c^{4}) = 0.745^{\,+0.090}_{\,-0.074} (\rm stat) {}^{\,+0.036}_{\,-0.036} (\rm syst) ~.
\end{equation} 

\noindent This measurement is consistent with the SM at the level of 2.6$\sigma$.
 It could be consistent with the $Z'$ interpretation of the anomaly in the \decay{\Bz}{\Kstarz\mumu} angular distribution if the $Z'$ couples preferentially to muons over electrons.

\section{Charm and top decays}

Up-type FCNC transitions happen at significantly lower rates than down-type transitions in the SM, due to the small size of the $b$-quark mass with respect to that of the $t$-quark. This makes the GIM cancellation much more effective in up-type transitions and leads to branching fraction predictions, for example,  at the level of $10^{-18}$ for the \decay{\Dz}{\mumu} compared to \decay{\Bs}{\mumu} at the level of $10^{-9}$.

The LHCb experiment has recently published several searches for rare charm decays~\cite{Aaij:2013cza, Aaij:2013sua, Aaij:2013sua}, improving existing limits from the B factories by about a factor of fifty. The results of these searches are

\begin{equation}
\begin{split}
\BF(\decay{\Dz}{\mumu}) & < 6.8\times 10^{-9} ~{\rm at~95\%~CL}  \\
\BF(\decay{D^{+}}{\pip\mumu}) & < 8.3\times 10^{-8}~{\rm at~95\%~CL}   \\ 
\BF(\decay{D_{s}^{+}}{\pip\mumu}) & < 4.1\times 10^{-7} ~{\rm at~95\%~CL}  \\ 
\BF(\decay{\Dz}{\pip\pim\mumu}) & < 5.5\times 10^{-7} ~{\rm at~90\%~CL}  \\
\end{split}
\end{equation} 

\noindent In all cases the measurements are several orders of magnitude above the SM predictions for rare FCNC charm decays, but probe branching fractions that are interesting for extensions of the SM~\cite{Burdman:2001tf}.  

The general purpose detectors are also able to look for FCNC top decays, which are exceedingly rare in the SM and interesting because the top quark is the heaviest of the SM particles. The CMS experiment has performed searches for decays \decay{t}{Z^0 + {\rm jet}}~\cite{Chatrchyan:2013nwa}, where the \decay{Z^0}{\ellell} and the other top quark is reconstructed through its dominant decay \decay{t}{W^+ b}. CMS sets a limit at the level 

\begin{equation}
\BF(\decay{t}{Z^0 + {\rm jet}}) < 5\times 10^{-4} {\rm ~at~95\% CL}~,
\end{equation}

\noindent improving on previous limits from the ATLAS experiment. Due to the large top mass, it is also interesting to study FCNC Higgs decays. These results are not reported here, but are described in Refs.~\cite{ATLAS:tHc, CMC:tHc}.

The ATLAS experiment has also performed searches for FCNC top couplings by looking at anomalous single top production through quark + gluon $\to$ top. ATLAS sets limits of~\cite{Aad:2012gd}

\begin{equation}
\begin{split}
\BF(t \to u g) & < 5.7 \times 10^{-5}  ~{\rm at~95\%~CL}  \\ 
\BF(t \to c g) & < 2.7 \times 10^{-4}  ~{\rm at~95\%~CL} ~. \\ 
\end{split} 
\end{equation}  

\section{Conclusions}

The excellent performance of the LHC and the LHC experiments, along with large production cross sections at $\sqrt{s}$ of $7$ and $8\tev$, has enabled unprecedentedly large  samples of top, beauty and charm decays to be reconstructed. Using these large samples the CMS and LHCb experiments have seen evidence of the very rare decay \decay{\Bs}{\mumu} for the first time. The LHCb experiment has also been able to demonstrate for the first time that photons produced in \decay{b}{s\gamma} transitions are polarised using \decay{\Bp}{\Kp\pim\pip\gamma} decays.  In the SM, this polarisation is almost purely left-handed. A more detailed understanding of the $\Kp\pim\pip$ system will be needed to confirm that this is true in data. Precise measurements of the angular distribution of \decay{\Bz}{\Kstarz\mumu} decays have also started to be made by ATLAS, CMS and LHCb. The most precise of these, by LHCb, shows a local discrepancy with the SM.  The data collected by the LHC experiments in 2012 will help to resolve the discrepancy, but more work may be needed from the theoretical side to understand the nature of any discrepancy.

\end{document}